\documentclass[twocolumn,prd,superscriptaddress,nofootinbib]{revtex4-1}

\usepackage{graphicx}
\usepackage{amsmath,bm}
\usepackage[usenames,dvipsnames]{xcolor}
\usepackage[normalem]{ulem}
\usepackage{url}
\usepackage{multirow}
\usepackage[colorlinks  = true,
            linkcolor   = NavyBlue,
            urlcolor    = NavyBlue,
            citecolor   = NavyBlue,
            anchorcolor = NavyBlue]{hyperref}

\def \us {ODDK22}

\def \prd {Phy. Rev. D}

\begin{document}

\title{New determination of the neutrino hadronic production cross sections \\ from GeV to beyond PeV energies}

\author{Luca Orusa}
\email{luca.orusa@princeton.edu}  
\affiliation{Department of Astrophysical Sciences, Princeton University, Princeton, NJ 08544, USA}
\affiliation{Department of Astronomy and Columbia Astrophysics Laboratory, Columbia University, New York, NY 10027, USA}

\author{Mattia Di Mauro}
\email{dimauro.mattia@gmail.com}
\affiliation{Istituto Nazionale di Fisica Nucleare, via P. Giuria, 1, 10125 Torino, Italy}

\author{Fiorenza Donato}
\email{donato@to.infn.it}
\affiliation{Department of Physics, University of Torino, via P. Giuria, 1, 10125 Torino, Italy}
\affiliation{Istituto Nazionale di Fisica Nucleare, via P. Giuria, 1, 10125 Torino, Italy}

\begin{abstract}
\noindent
The flux of astrophysical neutrinos is now measured with unprecedented accuracy and over several decades of energy
spectrum. Their origin traces back to hadronic collisions between protons and nuclei in the cosmic rays with hydrogen and helium in the target gas. To accurately interpret the data, a precise determination of the underlying cross sections is therefore mandatory. 
We present a new evaluation of the neutrino production cross section from $p+p$ collisions, building on our previous analysis of the production cross section for $\pi^\pm$, $K^\pm$, and minor baryonic and mesonic channels.  Cross sections for scatterings involving nuclei heavier than protons are also derived. 
The novelty of our approach is the analytical description of the Lorentz invariant cross section $\sigma_{\rm inv}$ and the fit of the model to the available accelerator data. 
We work with neutrino energies from $10$~GeV to $10^7$~GeV and, correspondingly, to incident proton (nuclei) energies from $10$~GeV to $10^9$~GeV (GeV/n). 
We obtain the total differential cross section $d\sigma(p+p\rightarrow \nu+X)/dE_{\nu}$ as a function of neutrino and proton energies, with an estimated uncertainty of 5\% for neutrino energies below 100 GeV, increasing to 10\% above TeV energies. Predictions are given for 
$\nu_e, \nu_\mu, \bar{\nu_e}$ and $\bar{\nu_\mu}$. 
A comparison with state-of-the-art cross sections, all relying on Monte Carlo generators, is also presented. 
To facilitate the use by the community, we provide numerical tables and a script for accessing our energy-differential cross sections.
\end{abstract}

\maketitle

\section{Introduction}

The main origin of astrophysical neutrinos is intimately connected to hadronic processes. Being electrically neutral, neutrinos cannot be directly accelerated by electric fields; instead, they are produced in hadronic interactions when energetic particles collide with ambient matter. Such processes occur both in extragalactic sources, such as blazars, particularly in their flaring activity periods \cite{IceCube:2018dnn}, and star forming galaxies, for example from NGC 1068 \cite{IceCube2022}.
The IceCube Collaboration has measured a neutrino diffuse flux of astrophysical origin in the TeV-PeV energy range and with a spectral shape compatible with a power-law of index of about $2.5$ (see, e.g., \cite{IceCube:2013low,IceCube:2020acn,IceCube:2020wum,IceCube:2024fxo}). On the other hand the detection of a 120 PeV event by KM3NeT \cite{KM3NeT:2025npi} has further extended the frontier of neutrino physics.
Recently, the IceCube Collaboration has found strong evidence for a Galactic plane emission, which is consistent with the diffuse neutrino flux produced from interactions of cosmic rays (CRs)—predominantly protons and helium nuclei—with the interstellar medium (ISM) via hadronic collisions \cite{IceCube:2023ame}, even if a part of the events could arise from a population of unresolved point sources. 
Much like the diffuse emission of high-energy photons, neutrinos can yield valuable information on the spatial and spectral distribution of Galactic CRs throughout the Milky Way \cite{Schwefer_2023}. 
In fact, at variance with $\gamma$ rays, which can originate from both hadronic and leptonic processes, the diffuse neutrino flux provides a cleaner tracer of hadronic interactions of CRs. 

Any prediction of the neutrino emission, whether Galactic or from extragalactic sources, depends on several factors, including the spectra of incident CRs, the choice of target gas maps, and the hadronic interaction cross sections. In this work, we focus specifically on improving the estimation of the latter, presenting a new model for neutrino production cross sections, building on our previous analyses reported in \cite{Orusa:2022pvp, Orusa_2023}. Recently, the importance of cross sections in astroparticle physics has been extensively discussed in \cite{Maurin:2025gsz}.
The most direct applications of our new cross sections span processes of astrophysical neutrino production, both in diffuse or point source emissions. The applicability of our results can nonetheless be wider.

For example, the inclusive neutrino production cross sections we provide for $p+p\to \text{hadrons}\to \nu$ could be applicable to proton beam-dump and fixed-target searches for light dark matter, where neutrino-induced neutral-current interactions constitute leading backgrounds. In contrast, for central missing-transverse-momentum (MET) searches at the LHC (monojet, $\gamma+$MET, jets+MET), the dominant Standard Model backgrounds arise from electroweak processes—chiefly $pp\to Z(\nu\bar\nu)+\text{jets}$ (irreducible) and $pp\to W(\ell\nu)+\text{jets}$ with an undetected lepton—along with top and diboson production \cite{ATLAS:2102.10874,CMS:2107.13021}, which are not the main focus of this paper. Neutrinos from pion and kaon decays produced in generic hadronic activity are predominantly forward and low transverse momentum, and therefore contribute negligibly to large MET in the central detectors. Our cross sections could therefore be directly applied to predict neutrino fluxes in forward LHC detectors and to evaluate neutrino backgrounds in proton beam-dump configurations, after appropriate generalization to $p+A$ collisions and inclusion of geometry and shielding effects.
\cite{FASER:2023PRL,SNDLHC:2023PRL,DarkQuest:2022Snowmass,SHiP:2022EPJC}.

The diffuse hadronic neutrino emission depends on the CR fluxes, the density of the ISM, and the inelastic production cross section $\sigma(p + p \rightarrow \nu + X)$ (and analogously for heavier nuclear components in both CRs and the ISM). Local CR fluxes are measured with high precision by AMS-02~\cite{AMS:2021nhj}, DAMPE~\cite{Dampe_2019}, and CALET~\cite{Adriani_2022}, and the ISM density in our nearby Galactic environment, within a few kiloparsecs, is also relatively well constrained~\cite{Widmark:2022qgx}. However, beyond the solar neighborhood, the situation becomes more uncertain: CR fluxes must be extrapolated from local measurements, making predictions  model-dependent. Likewise, determining the distribution of gas in more distant regions of the Galaxy is considerably more challenging~\cite{Pohl:2007dz, Mertsch:2022oee}.

A key ingredient for accurately predicting the hadronic diffuse $\nu$ emission is the inclusive neutrino production cross section $\sigma(p + p \rightarrow \nu + X)$, and other reactions involving He and heavier nuclei. The standard approach to calculate these cross sections is to employ Monte Carlo event generators~\cite{Kamae:2006bf,Kachelriess:2019ifk,Bhatt_2020,Mazziotta_2016}. The most widely used model is based on a customized implementation of \textsc{Pythia}~6 by Kamae et al.~\cite{Kamae:2006bf}. For example this is the cross section theoretical framework employed by the IceCube Collaboration in \cite{IceCube:2023ame}. More recent results include those from \textsc{AAfrag}~\cite{Kachelriess:2019ifk}, based on the QGSJET-II-04m event generator, and from Bhatt et al.~\cite{Bhatt_2020}, based on the DPMJET-III-19.1 event generator.
Significant discrepancies between Monte Carlo simulations and experimental data have been pointed out—for example, in the production cross sections of $\bar{p}$~\cite{Kachelriess:2015wpa,Kachelriess:2019taq}, in $e^{\pm}$~\cite{Orusa:2022pvp} (hereafter \us), and in $\gamma$ rays~\cite{Orusa_2023}. Moreover, as shown in Refs.~\cite{Koldobskiy:2021nld,dorner_25}, the predicted production cross sections of all $\nu$ flavors can differ by up to a factor of $\mathcal{O}(2)$ depending on the Monte Carlo generator used. These large variations underscore the need for improved modeling of neutrino production cross sections.

In this paper, we present a new accurate model that relies primarily on an analytic prescription fitted to the available experimental data, with the aim of determining their correct dependence on kinematic variables and of robustly quantifying the modeling uncertainties. The dominant production channel for $\nu_e$, $\nu_\mu$, $\bar{\nu_e}$ and $\bar{\nu_\mu}$ is the decay of $\pi^\pm$ mesons, for which we build upon the analysis performed in \us. In addition, we carefully model the production cross sections of $K$ mesons and $\Lambda$ baryons, which contribute as well to the $\nu$ flux through their decay into $\pi^\pm$. Our strategy closely follows the approach adopted in \us\ for deriving cross sections relevant to the secondary production of CR electrons and positrons. While the parameterization of the individual channel cross sections were derived in \us\, the main innovation of this paper lies in the integration of these channels into a complete, data-driven model for the inclusive neutrino production cross section, along with a systematic quantification of the associated uncertainties.

The remainder of this paper is organized as follows. In Sec.~\ref{sec:emissivity}, we outline the theoretical framework used to derive the observed $\nu$ flux from hadronic production cross sections. Section~\ref{sec:pi_charged} focuses on the analytical modeling of the $\pi^\pm$ production cross section, the dominant channel for neutrino generation. In Sec.~\ref{sec:other_channels}, we estimate contributions from additional production channels and from interactions involving nuclei. Our results are presented in Sec.~\ref{sec:results}, followed by our conclusions in Sec.~\ref{sec:conclusions}.

\begin{figure*}[t]
  \centering {
    \includegraphics[width=0.95\textwidth]{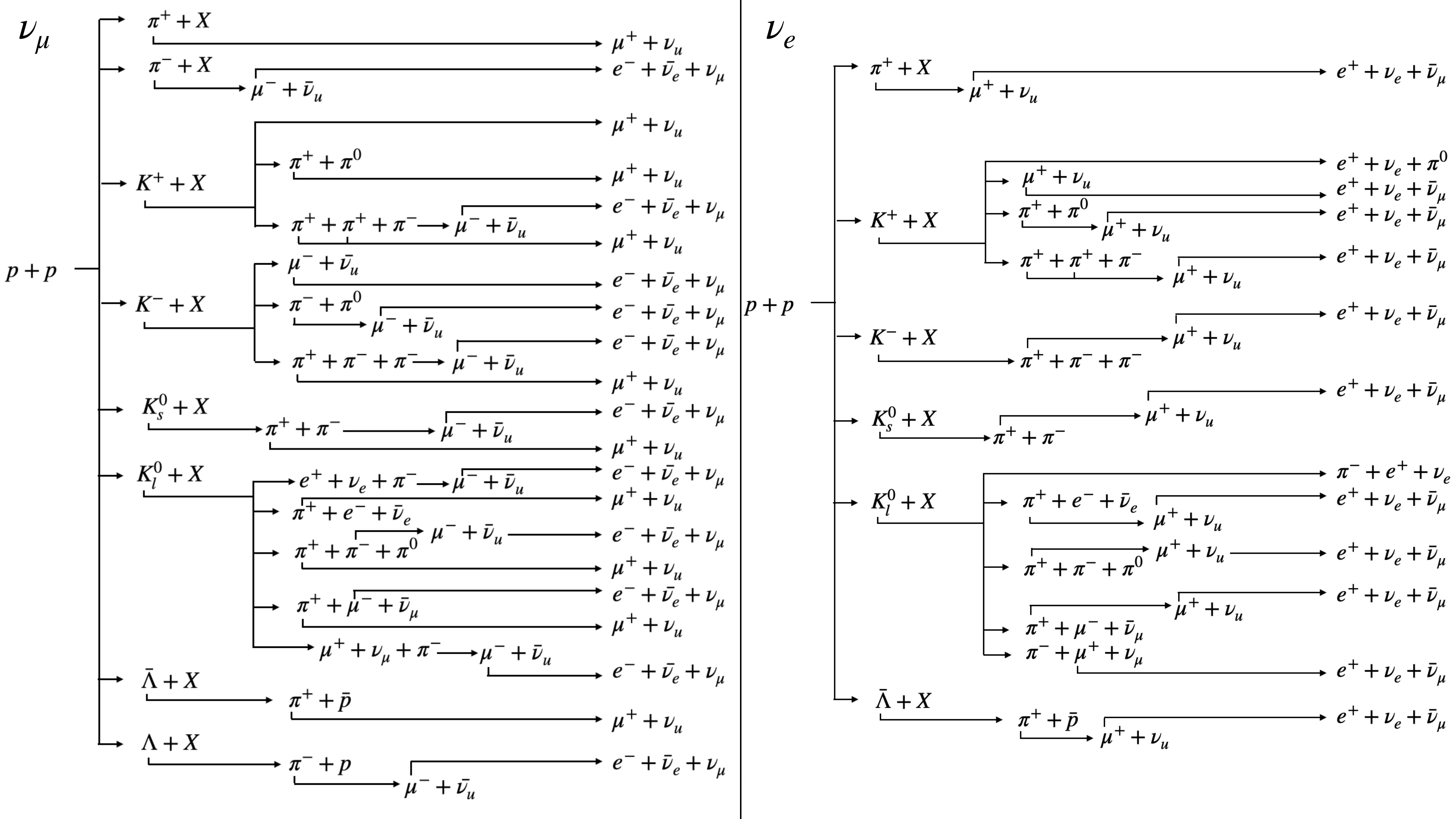}
  }
  \caption{
             This diagram shows the $\nu_\mu$ and $\nu_e$ production channels from $p+p$ collisions considered in our analysis. The same scheme applies to $\bar{\nu}_\mu$ and $\bar{\nu}_e$ under charge conjugation (with the exception of the initial $p+p$ state). Only channels contributing at least 0.5\% of the total yield are reported (see main text for details).
             }
             \label{fig:nu_production_channels}
\end{figure*}

\section{Neutrino production cross sections and  $\nu$ emissivity}
\label{sec:emissivity}
The original results of this paper relate to a new determination of the $\nu$ production cross section from hadronic collisions. 
Throughout this section, we refer generically to the neutrino flavor as $\nu$, while the contributions from individual flavors will be discussed in more detail in the following sections.
Given the relevance of these processes in the computation of the flux from astrophysical sources and environments, we also provide results for the $\nu$ emissivity, which is the fundamental brick in the flux determination 
in astrophysical contexts. 
The flux observed at Earth, $\phi_\nu$, is given by the integral along the line-of-sight (l.o.s.) of the $\nu$ emissivity, $\epsilon_{i,j}$, summed over all combinations of incoming CR species $i$ interacting on the ISM components $j$:

\begin{eqnarray}
    \label{eq:gamma_ray-flux}
    \frac{d^2 \phi_\nu}{d \Omega d E_\nu} (E_\nu, l, b) 
    = 
    \sum\limits_{ij}  \int\limits_{\text{l.o.s.}} \!\!\!\!\! \;d \ell \;\; \epsilon_{ij}\Big(\Vec{x}(\ell, l,b), E_\nu\Big) \, .
\end{eqnarray}
Here,  $\Vec{x}$ is the source position, $\ell$ the distance along the l.o.s., while $l$ and $b$ denote the Galactic longitude and latitude, respectively. The flux is expressed as a differential quantity in $\nu$ energy, $E_\nu$, and solid angle, $\Omega$. The total flux is computed  by integrating over the full solid angle.

The emissivity at a given Galactic location $\vec{x}$ is obtained by convolving the local CR flux at a given position and kinetic energy $T_i$, $\phi_i(\Vec{x},T_i)$, and the ISM density $n_{\mathrm{ISM},j}(\Vec{x})$, with the differential cross section for $\nu$ production, $d\sigma_{ij}(T_i,E_{\nu})/dE_\nu$ for the reaction $i + j \rightarrow \nu + X$:

\begin{equation}
\label{eq:source_term}
\epsilon_{ij}( \Vec{x}, E_\nu) = n_{\mathrm{ISM},j}(\Vec{x})
\int dT_i \, \phi_i(\Vec{x},T_i)\frac{d\sigma_{ij}}{d E_{\nu}}(T_i,E_{\nu}).
\end{equation}
In general, the emissivity varies across the Galaxy, reflecting spatial variations in both the CR flux and the ISM density, and the properties of neutral particles, which are not deflected during propagation. 

We deal here with neutrinos produced in hadronic scatterings, mainly $p+p$, therefore arising from the decays of mesons and baryons. In Fig.~\ref{fig:nu_production_channels}, we illustrate the relevant production channels for $\nu_\mu$ and $\nu_e$ that have been included in the present analysis. The corresponding antineutrino channels can be obtained by replacing each particle with its antiparticle.
The dominant contribution comes from charged pions, which subsequently decay into neutrinos. This channel is examined in detail in Sec.~\ref{sec:pi_charged}. The additional channels shown in Fig.~\ref{fig:nu_production_channels} are discussed in Sec.~\ref{sec:other_channels}. Channels contributing less than $0.5\%$ to the total $\nu$ yield are omitted in the figure and neglected in our analysis. Although some of these minor channels are poorly constrained and difficult to quantify precisely, we estimate  
their combined contribution to be at the level of $\sim 1\%$, which remains below the overall uncertainty budget.

In the following of this section we specify formulae for the pion production case, 
but they are formally identical 
for any other meson or baryon directly produced in the $p+p$ collision, as listed in Fig.~\ref{fig:nu_production_channels}. 
The $\nu$ production cross section is obtained from the charged-pion one via the convolution:
\begin{equation}
\label{eq:convolution}
\frac{d\sigma_{ij}}{d E_{\nu}}(T_i,E_{\nu})=
\int d T_{\pi^\pm} \, \frac{d\sigma_{ij}}{dT_{\pi^\pm}}(T_i,T_{\pi^\pm})
\; P(T_{\pi^\pm}, E_{\nu}) ,
\end{equation}
where $T_{\pi^\pm}$ denotes the kinetic energy of the pion that decays into a $\nu$ with energy $E_{\nu}$. The probability distribution function $P(T_{\pi^\pm}, E_{\nu})$ for this decay can be computed analytically, and is adopted from Ref.~\cite{kelner+06}.

The fully differential production cross section can be expressed in the Lorentz-invariant form:
\begin{equation}
\sigma^{(ij)}_{\rm inv} = E_{\pi^{\pm}} \frac{d^3 \sigma_{ij}}{dp_{\pi^{\pm}}^3},
\label{eq:invariant}
\end{equation}
Here, $E_{\pi^{\pm}}$ denotes the total energy and $p_{\pi^{\pm}}$ the momentum of the charged pion. This cross section depends on three kinematic variables: the center-of-mass (CM) energy $\sqrt{s}$, the transverse momentum $p_T$, and the radial scaling variable $x_R$, defined as the ratio of the pion energy to its maximum possible value in the center-of-mass frame,  
$x_R = E_{\pi^{\pm}}^\ast/E_{\pi^{\pm}}^{\max\ast} \, .$
The energy-differential cross section in Eq.~\eqref{eq:convolution} is obtained by first transforming the kinematic variables from the CM frame to the fixed-target (LAB) frame, and then integrating over the solid angle $\Omega$:
\begin{eqnarray}
    \frac{d\sigma_{ij}}{d T_{\pi^{\pm}}}(T_i,T_{\pi^{\pm}}) &=& p_{\pi^{\pm}}\int d \Omega \; \sigma_{ {\rm inv}}^{(ij)}(T_i,T_{\pi^{\pm}},\theta) \\
    &=& 2\pi p_{\pi^{\pm}} \int^{+1}_{-1} d(\cos{\theta}) \; \sigma_{ {\rm inv}}^{(ij)}(T_i,T_{\pi^{\pm}},\theta)\,, \nonumber
    \label{eq:solid_int}
\end{eqnarray}
where $\theta$ is the angle between the incoming projectile and the outgoing $\pi^\pm$ in the LAB frame. In the following, we detail the calculation of the $\nu$ production cross sections from $\pi^\pm$ decays, making use of the $\pi^\pm$ production cross sections derived in \us.

\section{$\nu$ from $p+p \rightarrow \pi^\pm + X$ collisions}
\label{sec:pi_charged}

\begin{figure*}[t]
  \centering {
    \includegraphics[width=0.99\textwidth]{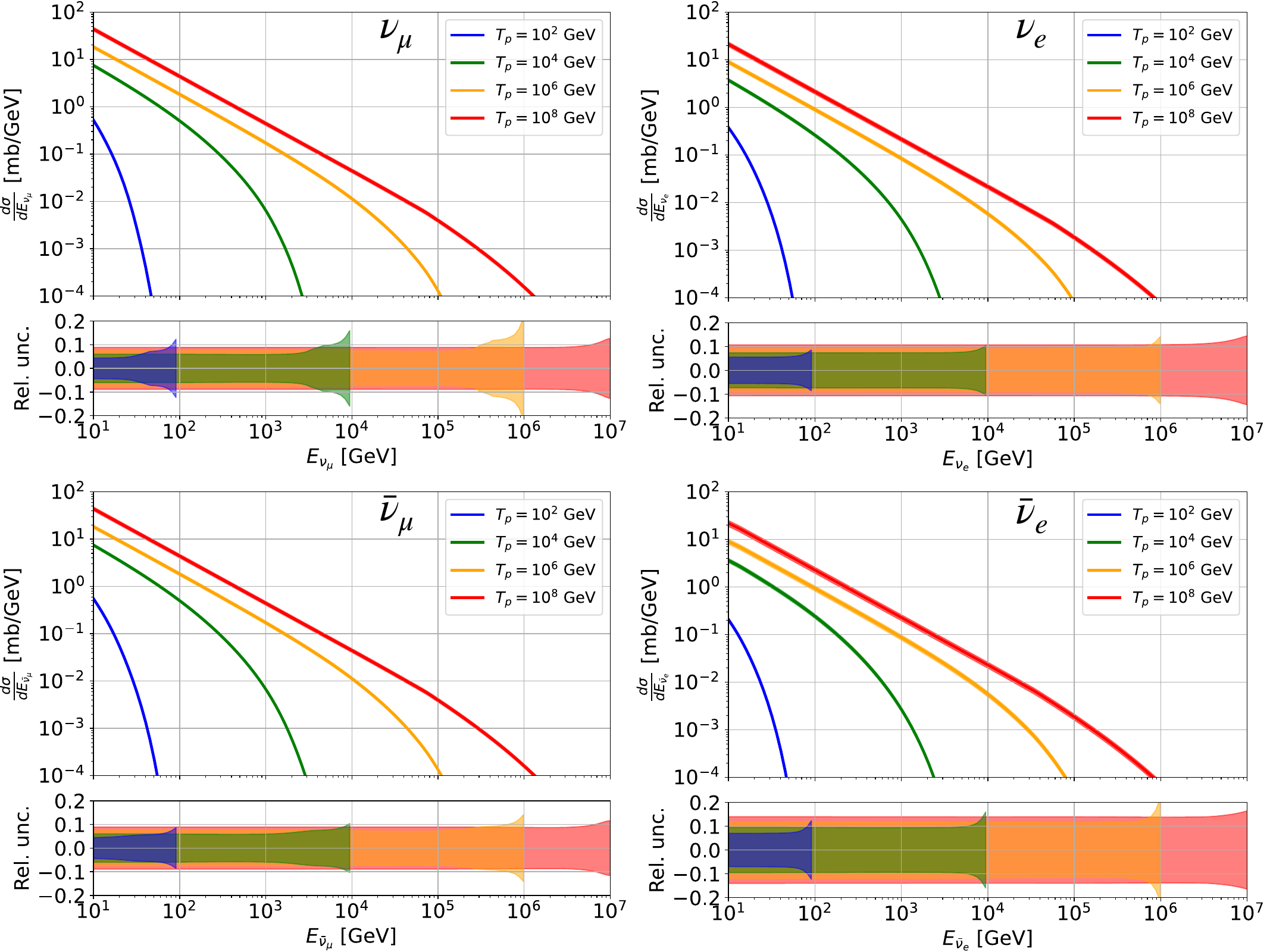}
    }
  \caption{
            Differential cross section for the production of $\nu_\mu$, $\nu_e$ and relative antiparticles  from $\pi^\pm$ in $p+p$ collisions, computed for different incident kinetic energies of the proton $T_p$.}
\label{fig:pion_contribution}
\end{figure*}

The $\pi^\pm$ channel for $\nu$ production 
 corresponds to the first two rows in Fig. \ref{fig:nu_production_channels}(left) for $\nu_\mu$ and to the first row in Fig. \ref{fig:nu_production_channels}(right) for $\nu_e$.
 Given its relevance, it is crucial to obtain precise data covering a wide kinematic phase space for the reaction $p+p \rightarrow \pi^\pm + X$. 
As discussed in \us, measurements of $\sigma_{\rm inv}(p+p \rightarrow \pi^\pm + X)$
have been collected by various accelerator and collider experiments, spanning large portions of the kinematic phase space. In particular, NA49~\cite{2005_NA49} and NA61~\cite{Aduszkiewicz:2017sei} provide data at low $\sqrt{s}$, while ALICE~\cite{2011_ALICE} and CMS~\cite{2012_CMS,2017_CMS} cover the high-$\sqrt{s}$ regime.  
In this work we do not perform a new fit to the particle production data (since no new data have appeared); instead, we use the results for $\sigma_{\rm inv}$ obtained in \us\ to start the calculation of $\nu$ yields. The expressions for $\sigma_{\rm inv}$ are given in Eqs.~(7)–(10)  and the parameters for $\pi^\pm$ are listed in Table~II of \us\ and in the Appendix \ref{appendix}. We remark that the fit performed in \us\ is based on an analytical formula for $\sigma_{\rm inv}$, directly calibrated on the data and their uncertainties.
The highest projectile energy on which we tested our model is $T_p = 9 \times 10^7$~GeV, corresponding to  $\sqrt{s} = 13$~TeV pion data from CMS~\cite{2017_CMS}. Beyond this limit, our model can only be extrapolated. 
However, care must be taken when treating purely statistical uncertainties at high $\sqrt{s}$. As explained in ODDK22, our model of $\sigma_{\rm inv}$ relies on a fit on NA49 and NA61 $\pi^\pm$ data at $\sqrt{s} = 17.3$ GeV, which provide broad coverage in both $p_T$ and $x_R$, thereby constraining in a solid way the dependence of the model on these variables. The scaling of the cross section at higher energies was instead determined by fits to CMS and ALICE data at $\sqrt{s} > 900$ GeV (corresponding to $T_p=4.3 \times 10^5$ GeV in the LAB frame). These measurements, however, were performed at mid-rapidity and cover only a limited range in $p_T$, with no data available at forward rapidities. Consequently, our model implies an extrapolation of the cross section in the forward-rapidity region at high $\sqrt{s}$. To quantify the impact of this incomplete phase-space coverage, we will compare our results with predictions from other state-of-the-art Monte Carlo models, thereby estimating the overall uncertainties across different cross-section parametrizations.
\\
\indent
In Fig.~\ref{fig:pion_contribution}, we show the differential cross section for the production of $\nu_\mu$, $\bar{\nu}_\mu$, $\nu_e$, and $\bar{\nu}_e$ from $\pi^\pm$ decays in $p+p$ collisions.  
Results are provided for different incident kinetic proton energies $T_p$ as a function of $E_\nu$, for all $\nu$ flavors. The uncertainties range from $5\%$ to $20\%$ over most of the energy range, except for $E_{\nu}$ values close to $T_p$, where statistical errors increase.  
The most accurate predictions are obtained around $T_p \sim 100~\mathrm{GeV}$, corresponding to the energy range covered by NA49 and NA61 data for $\pi^\pm$ production.  
Both $\nu_\mu$ and $\bar{\nu}_\mu$ receive contributions from $\pi^+$ and $\pi^-$ decays, as both pion charges can produce a $\nu_\mu$.  
In contrast, $\nu_e$ and $\bar{\nu}_e$ are primarily produced by $\pi^+$ and $\pi^-$ decays, respectively.  
As previously noted in \us, the modeling of $\pi^-$ production carries slightly larger uncertainties due to less precise experimental data, leading to $\bar{\nu}_e$ having the largest model uncertainty among all $\nu$ flavors.

\section{Contribution from other production channels and from nuclei}
\label{sec:other_channels}

In this section, we present our model for $\nu$ production from all the production channels other than the 
$\pi^\pm$ ones discussed in Sect. \ref{sec:pi_charged}.
We also remind the reader of the tools for treating scatterings involving nuclei heavier than hydrogen. 

The relevant decay channels contributing to $\nu_\mu$ production, for which we have an analytical model calibrated on the available data, along with their corresponding branching ratios (in parentheses), are:

\begin{itemize}
\item $K^+ \rightarrow \mu^+ \nu_\mu$ (63.6\%)
\item $K^+ \rightarrow \pi^+ \pi^0$ (20.7\%)
\item $K^+ \rightarrow \pi^+ \pi^+ \pi^-$ (5.58\%)
\item $K^- \rightarrow \mu^- \bar{\nu}_\mu$ (63.6\%)
\item $K^- \rightarrow \pi^- \pi^0$ (20.7\%)
\item $K^- \rightarrow \pi^- \pi^- \pi^+$ (5.58\%)
\item $K^0_S \rightarrow \pi^+ \pi^-$ (69.2\%)
\item $\Lambda \rightarrow p \pi^-$ (63.9\%)
\end{itemize}
The channels contributing to $\nu_e$ production include:
\begin{itemize}
\item $K^+ \rightarrow \mu^+ \nu_\mu$ (63.6\%)
\item $K^+ \rightarrow \pi^+ \pi^0$ (20.7\%)
\item $K^+ \rightarrow \pi^+ \pi^+ \pi^-$ (5.58\%)
\item $K^+ \rightarrow \pi^0 e^+ \nu_e$ (5.1\%)
\item $K^- \rightarrow \pi^- \pi^- \pi^+$ (5.58\%)
\item $K^0_S \rightarrow \pi^+ \pi^-$ (69.2\%)
\end{itemize}
The corresponding decay channels of the antiparticles contribute to the production of $\bar{\nu}_\mu$ and $\bar{\nu}_e$.  

We include the production cross sections derived in \us. The $\nu$ spectra are calculated assuming that $\pi^\pm$ are produced via either a two-body or a three-body decay. In particular, for three-body decays we follow the approach in \us, assuming that each of the three particles carries one-third of the parent's energy.  

The decay time of the $K^0_L$ meson is $5.1 \times 10^{-8}$~s, about 600 times longer than that of $K^0_S$, which makes detecting $K^0_L$ particles in accelerator experiments particularly challenging. Moreover, the $K^0_L$ has different decay channels and branching ratios compared to $K^0_S$, and contributes to $\nu_\mu$ production through:

\begin{itemize}
    \item $K^0_L \rightarrow \pi^{\pm} e^{\mp} \nu_e$ ($B_r=40.6\%$),
    \item $K^0_L \rightarrow \pi^{\pm} \mu^{\mp} \nu_{\mu}$ ($B_r=27.0\%$),
    \item $K^0_L \rightarrow \pi^{+} \pi^{-} \pi^{0}$ ($B_r=12.5\%$).
\end{itemize}

The lack of experimental data prevents an independent parametrization of the production cross section.  
We therefore use the \textsc{Pythia} event generator to compare the $p_T$ and $x_F$ dependence of the final $\nu$ spectra from $K^0_S$ and $K^0_L$ decays, where $x_F=2 p_L/\sqrt{s}$ and $p_L$ is the particle longitudinal momentum.  
We find that the $p_T$ and $x_F$ distributions for $\nu_\mu$ production are very similar for $K^0_L$ and $K^0_S$, differing only in normalization.  
Considering the different branching ratios for $K^0_S$ and 
$K^0_L$ and their decay chains, one finds that 
the $\nu_\mu$ yield from $K^0_L$ is about $94\%$ of the $K^0_S$ one. 
Similarly, for $\nu_e$ the yield from $K^0_L$ is 1.16 times than from $K^0_S$. 
In the following, we therefore assume that the $\nu_\mu$ production cross section from $K^0_L$ can be obtained by rescaling that from $K^0_S$ by a factor $0.94$, and  the $\nu_e$ one by a factor $1.16$.  
By charge symmetry, the same scaling factors are applied to $\bar{\nu}_\mu$ and $\bar{\nu}_e$.  
No additional uncertainty is assigned to these numbers, as they follow directly from the well-measured branching ratios of $K^0_L$ and $K^0_S$ decays into pions.  
We then assign to the $K^0_L$ channel the same uncertainty of the $K^0_S$ one.

\subsection{Subdominant channels}
\label{sec:eplus_otherchannels}

\begin{figure*}[t]
    \includegraphics[width=0.99\textwidth]{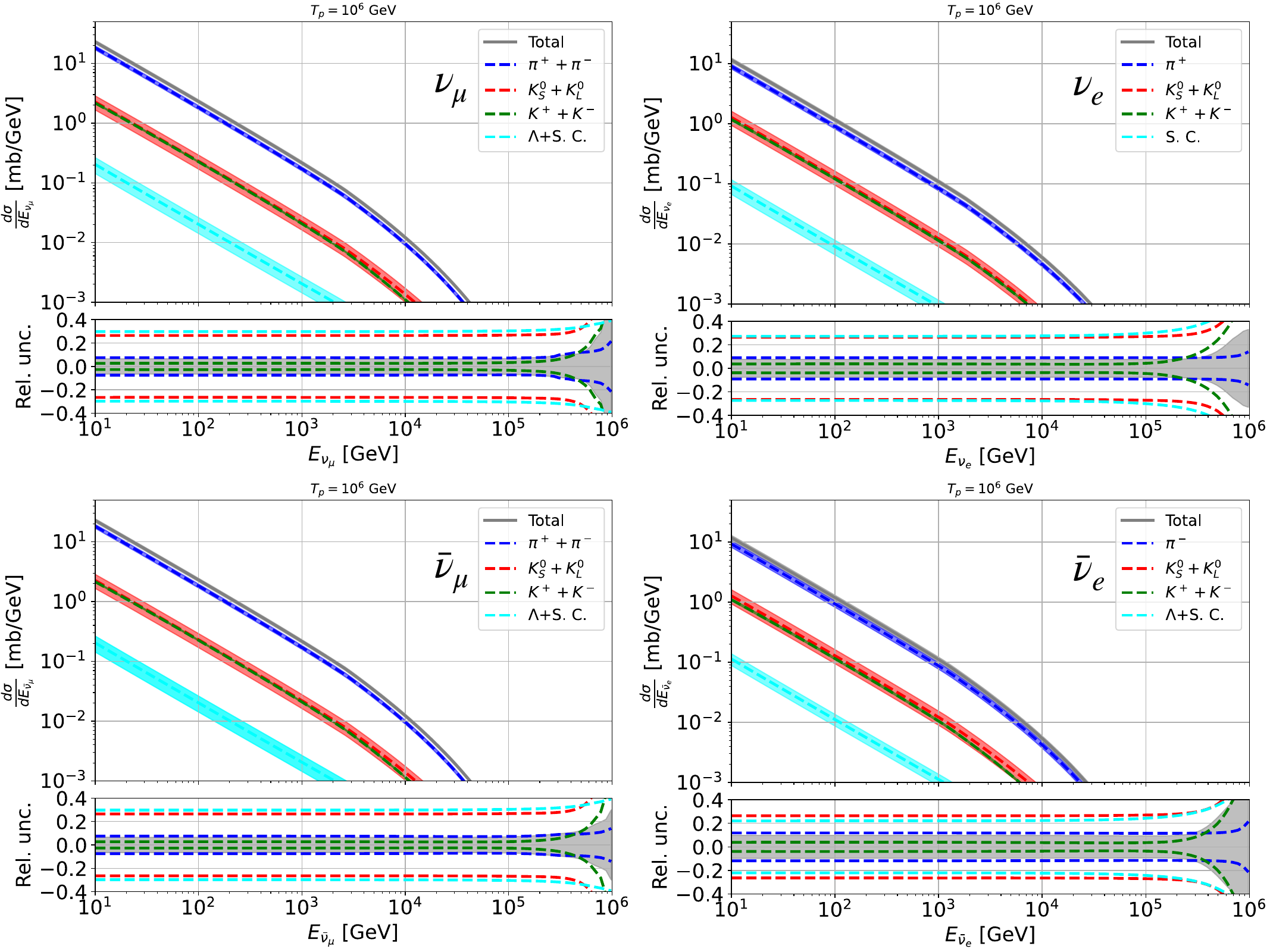}
    \caption{Differential cross section for the inclusive production of the different $\nu$ flavors and relative antiparticles in $p+p$ collisions, derived from fits to the data as described in Secs. \ref{sec:pi_charged} and \ref{sec:other_channels}. We plot separate production of $\pi^\pm$, $K^\pm$, $K_0^S$, $K_0^L$, and $\Lambda$ plus subdominant channels (S.C.), and their sum. Each plot is computed for incident proton energies $T_p=10^6$ GeV. The curves are displayed along with their 1$\sigma$ error band. At the bottom of each panel the 1$\sigma$ uncertainty band is displayed around the best fit individually for each contribution.} 
    \label{Fig:cross-final}
\end{figure*}
Other channels contribute only subdominantly to the yields of $\nu_\mu$, $\nu_e$, and their respective antiparticles. The $\bar{\Lambda}$, charged $\Sigma$, and $\Xi$ hyperons have typical lifetimes of order $10^{-10}$ s. Their contributions to pion production are usually not included since they decay outside the detector.  We therefore include them explicitly in our calculations, while neglecting other minor contributions.
Since no experimental data are available, in \us\  we estimated the contributions of the $\bar{\Lambda}$, $\Sigma$, and $\Xi$ baryons using the \textsc{Pythia} code~\cite{Sjostrand:2014zea}.
Specifically, we assume that the contributions of these minor channels to $\nu_\mu$ and $\nu_e$ are equal to that of $\Lambda$ to $\nu_\mu$ and $\bar{\nu}_e$, rescaled by a normalization factor $\mathcal{F}$ determined with Pythia. This factor accounts for the production multiplicities of the different hyperons and their branching ratios into channels that can decay into $\pi^\pm$. This approach is justified because we have an explicit model for the invariant cross section of $\Lambda$ production, and, given that the masses of $\bar{\Lambda}$, $\Sigma$, and $\Xi$ are similar or identical, we expect their cross sections to share comparable kinematic dependencies.
\begin{figure*}[t]
  \centering {
    \includegraphics[width=0.99\textwidth]{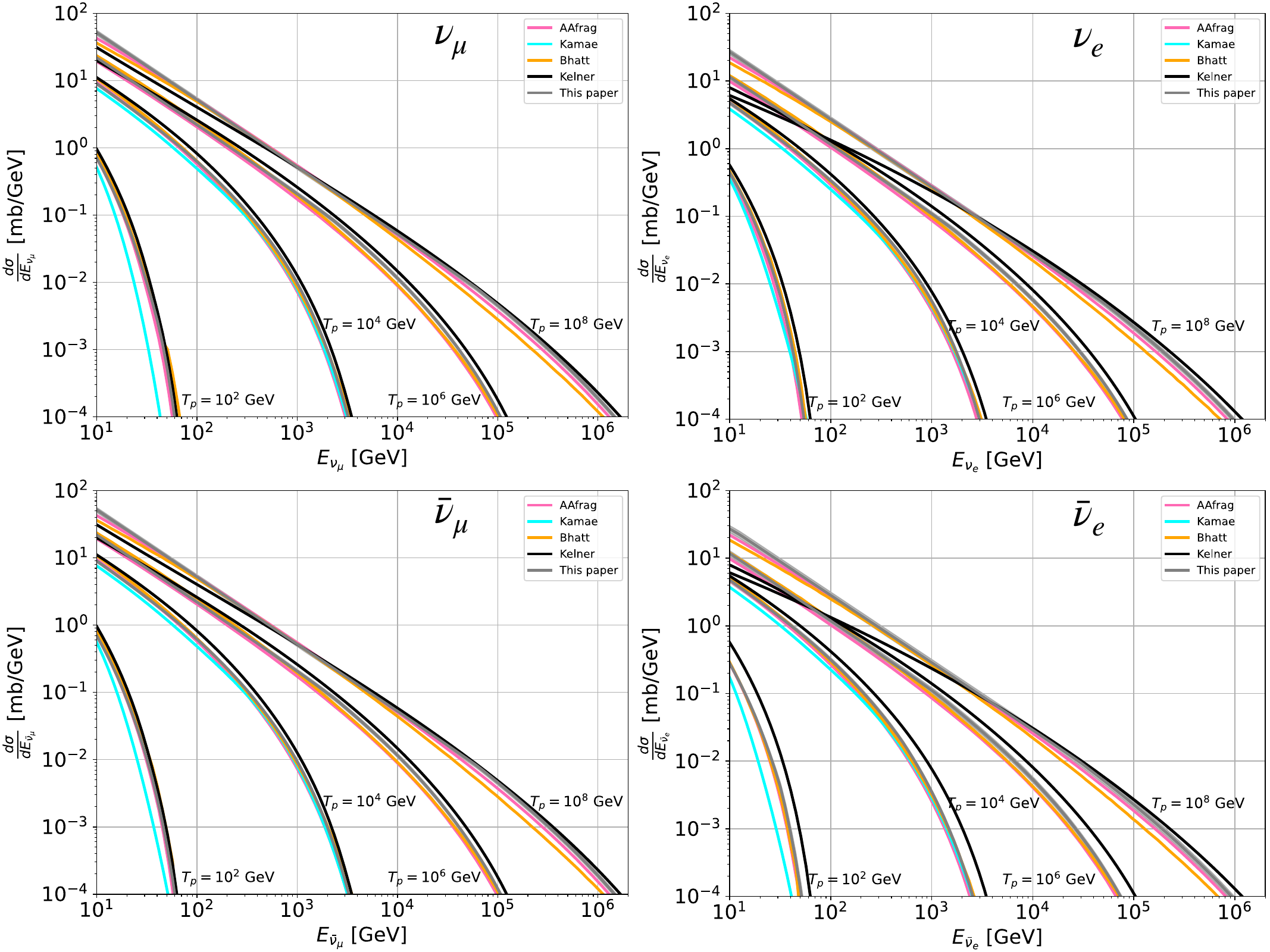}
    }
    \caption{Comparison among our differential cross sections and the one reported in \cite{Kamae:2006bf} (Kamae), \cite{Bhatt_2020} (Bhatt), \cite{Koldobskiy:2021nld} (AAfrag) and \cite{kelner+06} (Kelner) as a function of $E_\nu$, for incident proton energy $T_p =
    10^8, 10^6, 10^4$ and $ 10^2$ GeV.
   } 
    \label{Fig:qE_high_energy_comparison}
\end{figure*}

In Fig.~9 of \us, the $\mathcal{F}$ functions for $e^+$ and $e^-$ are shown. 
For $\nu_\mu$, the corresponding correction factor is given by the sum of the two functions 
calculated for $e^\pm$, since both $\pi^+$ and $\pi^-$ contribute to $\nu_\mu$ production. 
In contrast, $\nu_e$ production corresponds solely to the $\mathcal{F}_{e^+}$ function.

At low energies, $\mathcal{F}$ ranges from 20\% to 100\% for $\nu_\mu$ and from 10\% to 50\% for $\nu_e$, 
while at high energies it reaches up to 3 for $\nu_\mu$ and 2 for $\nu_e$. 
The details of the calculation are provided in \us. The correction factor can vary by as much as 40\% depending on the Monte Carlo setup; 
therefore, we associate a systematic uncertainty of 40\% with these channels across all energies.

\subsection{Production from heavier nuclei}
For the inclusion of scatterings involving nuclei heavier than hydrogen, either in the CRs or in the ISM, we closely follow the prescriptions derived in \us\ for $\pi^{\pm}$. Specifically, when a $\pi^\pm$ is produced in collisions between a projectile and a target nucleus with mass numbers $A_1$ and $A_2$, the $\sigma_{\rm inv}$ formulas reported in Eqs.~(7)--(10) of \us\ are modified according to the prescriptions in Eqs.~(25)--(27) of \us. 
The parameters in Eq.~(26) are taken from the columns 1 and 2 in Table V of \us, where the columns for $\pi^+$ ($\pi^-$) are used to correct $\sigma_{\rm inv}$.  The $K^\pm$ channel is modified analogously using columns 3 and 4 of Tab.~V in \us. For all other channels, we apply a correction function given by the average of the $K^+$ and $K^-$ cases.

\section{Results on the $\nu$ production cross section and emissivity}
\label{sec:results}

\begin{figure}[t]
  \centering {
    \includegraphics[width=0.49\textwidth]{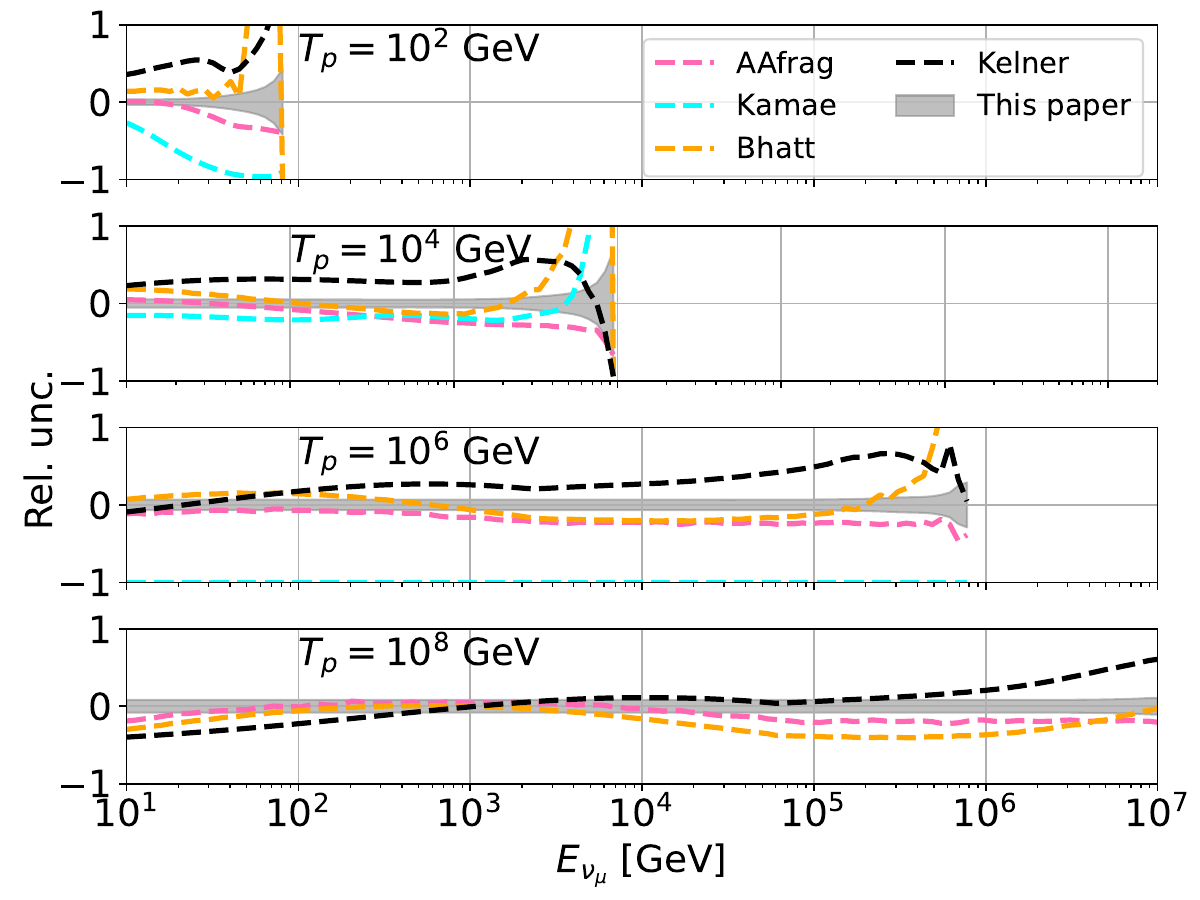}}
    \caption{Relative difference between our differential cross section and the one reported in \cite{Kamae:2006bf} (Kamae), \cite{Bhatt_2020} (Bhatt), \cite{Koldobskiy:2021nld} (AAfrag) and \cite{kelner+06} (Kelner), for incident proton energy $T_p =
    10^2, 10^4, 10^6$ and $ 10^8$ GeV for $\nu_\mu$. We also show the statistical uncertainty of our model as a gray band.
   }
    \label{Fig:qE_high_energy_relative}
\end{figure}

We have now all the tools to compute the total differential cross section $d\sigma/dE_{\nu}$ for the inclusive production of neutrinos in inelastic $p+p$ collisions. The result is obtained by summing all contributions from $\pi^\pm$ and the subdominant channels, as discussed in Secs~\ref{sec:pi_charged} and~\ref{sec:other_channels}. These cross sections constitute the main result of our paper and are shown in Fig.~\ref{Fig:cross-final} for a representative proton energy of $T_p = 10^6$~GeV, for all flavors of neutrinos and antineutrinos. 
The $\pi^\pm$ channels dominate at all $\nu$ energies and for all flavors. All kaons contribute about 20\% of the pions and 15\% of the total, while all other subdominant channels scrape the 1-2\% of the total. 
We have verified that these patterns are  maintained across different proton energies. 
The gray curve and shaded band show the total $d\sigma/dE_{\nu}$ and its $1\sigma$ uncertainty, respectively. The resulting uncertainty spans from $6\%$ to $20\%$ depending on $E_\nu$, and is primarily driven by the modeling of the $\pi^\pm$ cross section, with larger uncertainties for $\bar{\nu}_e$ due to the dominant production of $\pi^-$, as explained in Sec~\ref{sec:pi_charged}. We remind that our cross-section model is validated at high $\sqrt{s}$ by a fit to ALICE and CMS data probing the midrapidity region only. This ensures that we capture the correct dependence on $\sqrt{s}$ of the pion multiplicity, which is predominantly produced at midrapidity. However, the spectral shape of our cross section—depending on $p_T$ and $x_R$—relies on extrapolation at high $\sqrt{s}$. 

We compare our results, obtained from data-driven analytical approach, with state-of-the-art Monte Carlo codes.
In Fig.~\ref{Fig:qE_high_energy_comparison} we show $d\sigma/dE_{\nu}$ for different neutrino flavors and results from Monte Carlo codes as a function of neutrino energy and for various incident proton energies. 
Fig.~\ref{Fig:qE_high_energy_relative} (shown for $\nu_\mu$ only) zooms on  the relative difference between our model and the Monte Carlo predictions, with the statistical uncertainty of our model indicated by a gray band. As a general remark, our cross sections are in good agreement with the predictions by Bhatt, with discrepancies of the order of $10\%$, except for $E_\nu \sim T_p$, where the cross sections are nevertheless close to zero. The AAfrag model is also generally consistent with our results, particularly at low energies where deviations remain below $10\%$, while at higher energies differences increase to approximately $20\%$.

\begin{figure*}[t]
  \centering {
    \includegraphics[width=0.99\textwidth]{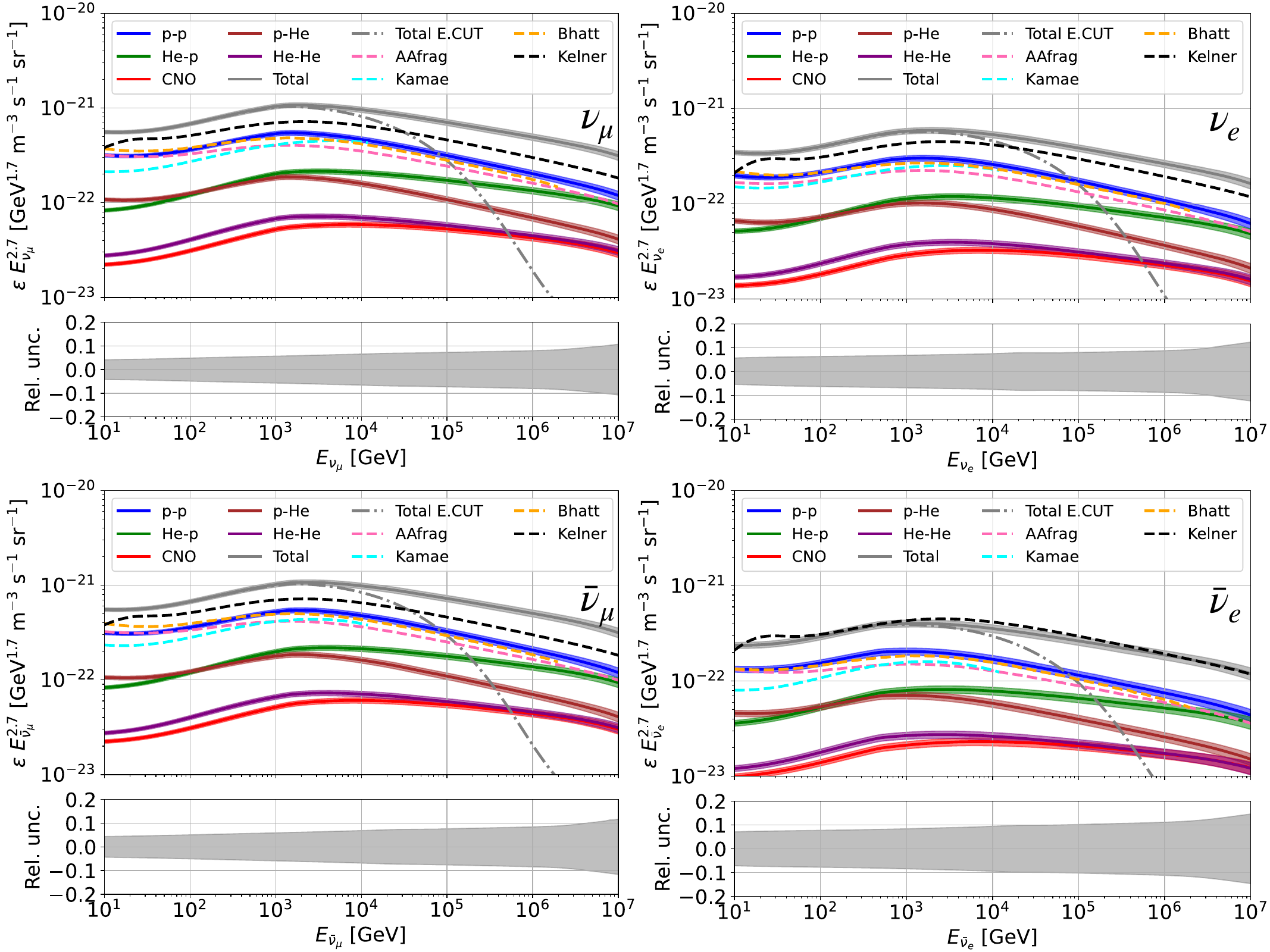}
    }
  \caption{The $\nu$ emissivity is computed for $p+p$, He$+p$, $p+$He, He+He and CNO$+p$ scatterings for the different neutrino and antineutrino flavors. The gray line is the sum of all contributions (see text for details). Each prediction is plotted with the relevant uncertainty due to the production cross sections derived in this paper. In each bottom panel, the relative uncertainty to the total $\epsilon(E_\nu)$ is displayed. For comparison, we show the results by \cite{Kamae:2006bf} (Kamae), \cite{Bhatt_2020} (Bhatt), \cite{Koldobskiy:2021nld} (AAfrag) and \cite{kelner+06} (Kelner) in the $p+p$ channel.\label{Fig:emissivity}}
\end{figure*}

In contrast, the Kamae et al.\ model shows deviations up to 50\%, particularly at low energies, and displays significantly different behaviours across the flavors, with the exception of $\nu_e$, which remains close to the other models. We do not report Kamae model for $T_p=10^6, 10^8$ GeV, since they do not provide the tables for these energies. 
This discrepancy could be particularly relevant for the observation of a significant Galactic diffuse neutrino emission made by the IceCube Neutrino Observatory
in \cite{IceCube:2023ame}, where their template model is based on Kamae's cross sections. 
In the end, the parametrization by Kelner \cite{kelner+06}—which does not provide separate cross sections for each neutrino and antineutrino species but instead gives averaged cross sections for $\nu_e$/$\bar{\nu}_e$ and $\nu_\mu$/$\bar{\nu}_\mu$—shows significant deviations from all other models, especially at high $T_p$. It predicts a noticeably harder dependence on $E_\nu$ compared to the other approaches, with particularly large discrepancies for $\nu_e$ and $\bar{\nu}_e$. In this case, using an average between the two species introduces substantial errors, especially at low energies where the $\pi^\pm$ production cross sections differ significantly.

In order to assess the impact of our new cross sections, we compute the emissivity for the different $\nu$ flavors and their corresponding antiparticles as in Eq.~\ref{eq:source_term}, assuming a constant $n_{\rm ISM}$ ($n_{\rm H} = 0.9~{\rm cm}^{-3}$ and $n_{\rm He} = 0.1~{\rm cm}^{-3}$) and incident CR spectra that are independent of Galactic position. The CR spectra are obtained from a fit to the AMS-02 \cite{AMS:2021nhj} and CALET data \cite{Adriani_2022,Adriani_2023}. In Fig.~\ref{Fig:emissivity}, we show $\epsilon(E_\nu)$ as a function of $E_\nu$ for $p+p$, He$+p$, $p+$He, He+He, and CNO$+p$ interactions, along with their sum. We plot two curves for the total emissivity, both shown in gray (solid and dot-dashed): one is obtained by extrapolating the high-energy behavior of the CR spectrum from fits to AMS-02 and CALET data (solid line), and the other assuming an exponential cutoff (dot-dashed) in the Galactic CR spectrum at 1 PeV for protons (scaling with the charge Z of each element, as the maximum energy is expected to be rigidity dependent). The presence or absence of such a cutoff at PeV energies leads to noticeable differences in the predicted emissivity already at $10^4$ GeV, consistent with the fact that predictions at a given $E_\nu$ depend in a non-trivial way on the integration over incident CR energies at least two orders of magnitude higher. 
Therefore, the accuracy of any neutrino production cross sections
depends on the proton energy or, equivalently, the beam energy at accelerators, well beyond the desired $E_\nu$. 

Each prediction provides the corresponding uncertainty from the production cross section. The relative uncertainty on the total $\epsilon(E_\nu)$ is shown in the bottom panels. As expected, the dominant contribution comes from $p+p$ interactions. However, scatterings involving He collectively yield a similar emissivity, with the He$+p$ channel becoming comparable to $p+p$ at very high energies for all $\nu$ flavors, due to the different high-energy behaviour hinted at by CALET measurements~\cite{Adriani_2023}. The decrease of the $p$/He CR ratio with energy explains why the predicted emissivities from $p+p$ and He$+p$ interactions converge at high energies, given that the cross section for the latter is roughly 3.15 times larger than that of the $p+p$ channel.
This plot is intended as an illustrative example to highlight the impact of our cross-section model once convolved with typical astrophysical source spectra. 

The statistical uncertainty on $\epsilon(E_\nu)$ due to hadronic production cross sections is approximately $5\%$ for $E_\nu \lesssim 10^3$~GeV, and increases to $>10\%$ for $E_\nu \gtrsim 10^6$~GeV. For comparison, we show the results in dashed lines from~\cite{Kamae:2006bf} (Kamae),~\cite{Koldobskiy:2021nld} (AAfrag), and~\cite{Bhatt_2020} (Bhatt) and \cite{kelner+06} (Kelner) for the $p+p$ channel. The models by Kamae and Bhatt do not provide full coverage across the proton and neutrino energy range required for this plot, and are therefore cut accordingly. 
There is excellent agreement at low energies ($E_\nu \lesssim 100$~GeV) among our model, AAfrag, and Bhatt. Overall, our results and Bhatt's (orange line) agree within $\sim10\%$, while AAfrag shows deviations up to $\sim20\%$. Conversely, Kamae exhibits significant discrepancies below $10^3$~GeV (except for $\nu_e$), while remaining broadly consistent with the other models between $10^3$~GeV and $10^4$~GeV. Kelner’s model shows significant discrepancies (e.g. up to 50\% or larger for $E_{\nu}>20$ GeV) at all energies and across all channels. The particularly large mismatch for the $\bar{\nu}_e$ channel (up to 100\% for $E_{\bar{\nu}_e}>20$ GeV) arises because the model uses an average of the $\pi^-$ and $\pi^+$ cross sections, which strongly overestimates the $\bar{\nu}_e$ yield. Overall, the Kelner parametrization substantially overestimates neutrino production in every channel, as already noted in \cite{Koldobskiy:2021nld}.
As a general trend, our new cross sections predict an emissivity that, at least for the dominant $p+p$ channel, is a bit higher that the Monte Carlo state of the art.

\section{Discussion and conclusions}
\label{sec:conclusions}

Hadronic collisions represent a primary mechanism for generating high-energy neutrinos in astrophysical environments. The neutrino production is dominated by the decay of charged pions, in turn produced by the inelastic scattering of CR nuclei with the target material. A precise modeling of the neutrino production cross section from hadronic interactions is crucial for interpreting current neutrino data.

In this paper, we present a new evaluation of the neutrino production cross section from $p+p$ collisions, building on our previous analysis of the production cross section for $\pi^\pm$, $K^\pm$, and minor baryonic and mesonic channels that contribute at least at the 0.5\% level. Cross sections for scatterings involving nuclei heavier than protons are also derived. The novelty of our approach, compared to the state of the art that relies on Monte Carlo predictions, is that we shape the $\sigma_{\rm inv}$ analytically, and directly fit the model to the available data. While the parameterization of the individual channel cross sections were derived in \us\, the main innovation of this paper lies in the integration of these channels into a complete, data-driven model for the inclusive neutrino production cross section, along with a systematic quantification of the associated uncertainties.
We work with neutrino energies from $10$~GeV to $10^7$~GeV, and correspondingly to incident proton (nuclei) energies from $10$~GeV to $10^9$~GeV (GeV/n). 
Our results include a realistic and conservative estimate of the uncertainties affecting the differential cross section $d\sigma/dE_{\nu}$, understood as the sum over all production channels. The uncertainty is estimated to be $5\%$ for $E_\nu \leq 1000$~GeV, increasing to $>10\%$ at higher energies in the TeV range and reaching 20\% at $E_\nu \simeq 10^7$ GeV. 

We compare our results to cross sections from state-of-the-art Monte Carlo codes, finding good agreement with \cite{Bhatt_2020} and \cite{Koldobskiy:2021nld}, but not with \cite{Kamae:2006bf}, 
which exhibits significant discrepancies with our predictions. 

To improve the accuracy of current results, new collider data are required. In particular, measurements of the Lorentz-invariant cross section for $\pi^\pm$ production at forward rapidities are essential. Recent efforts at the LHC to explore previously unmeasured regions of phase space -- such as those by LHCf \cite{Adriani_2016}, FASER \cite{Abraham_2024}, and SND@LHC \cite{Acampora_2024} -- could also be leveraged to provide measurements relevant to neutrino production, for reactions of interest in astrophysical studies \cite{Maurin:2025gsz}. It would also be important to perform similar measurements using a He target.
Our results could also be relevant for proton beam-dump and fixed-target searches for light dark matter, where neutrino-induced neutral-current interactions constitute leading backgrounds
\cite{FASER:2023PRL,SNDLHC:2023PRL,DarkQuest:2022Snowmass,SHiP:2022EPJC}.

We provide numerical tables for the energy-differential cross sections $d\sigma/dE_{\nu}$ as a function of $E_{\nu}$ from $10$~GeV to $10^7$~GeV, and for incident proton (nuclei) energies from $10$~GeV to $10^9$~GeV (GeV/n). A script to read the tables is also included. The material is available at \url{https://github.com/lucaorusa/neutrino_cross_section}.

\section*{Acknowledgments} 
LO thanks Julien Dörner for inspiring discussions. LO acknowledges the support of the Multimessenger Plasma Physics Center (MPPC), NSF grant PHY2206607. FD thanks the Department of Theoretical Physics of CERN, where part of this work for carried on. 
MDM and FD acknowledge support from the research grant {\sl TAsP (Theoretical Astroparticle Physics)} funded by Istituto Nazionale di Fisica Nucleare (INFN), MDM from the Italian Ministry of University and Research (MUR), PRIN 2022 ``EXSKALIBUR – Euclid-Cross-SKA: Likelihood Inference Building for Universe’s Research'', Grant No. 20222BBYB9, CUP I53D23000610 0006, and from the European Union -- Next Generation EU, 
and FD from the 
Research grants {\sl The Dark Universe: A Synergic Multimessenger Approach}, No.~2017X7X85K, funded by the {\sc Miur}. 

\bibliography{paper}

\appendix
\section{Lorentz invariant cross sections}\label{appendix}
For the sake of completeness, we report all the formulas taken from \us\ and employed 
in this paper.
\subsection{$\pi^\pm$ model}

The Lorentz invariant cross section for $\pi^\pm$ is given by:
\begin{equation}
   \sigma_{\rm inv}= \sigma_0 (s) \,  c_1 \,   \Big[F_p(s, p_T, x_R) + F_r(p_T, x_R)\Big]
    \, A(s),
   \label{eq:main_equation}
\end{equation}
where $\sigma_0 (s)$ is the total inelastic $p+p$ cross section taken from \us.  
The functional form of $F_p(p_T, x_R)$ is:
\begin{eqnarray}     
     \label{eq:function_prompt}
     F_p&&(s, p_T, x_R) = (1-x_R)^{c_2} 
        \exp(-c_3 \, x_R) \, p_T^{c_4}  \\ \nonumber
      &&\times \exp\left[ -c_5 \sqrt{s/s_0}^{\;c_6} 
      \left(\sqrt{p_T^2 + m_\pi^2}-m_{\pi}\right)^{c_7 \sqrt{s/s_0}^{\;c_6}} \right] \;, 
\end{eqnarray}
where $\sqrt{s_0}=17.3$ GeV. The model parameters $c_i$ are reported in Table II of \us.

On the other hand, the functional form of $F_r(p_T, x_R)$ reads:
\begin{eqnarray}
  \label{eq:fr}
  F_r&&(p_T, x_R) = (1-x_R)^{c_{8}}  \\ \nonumber
  &&\times \exp\left[ -c_{9}\,p_T - \left(\frac{|p_T-c_{10}|}{c_{11}}\right)^{c_{12}}\right]  \\ \nonumber
  &&\times \biggl[c_{13}\exp(-c_{14}\,p_T^{c_{15}} x_R) +  \\ \nonumber
  && \qquad + c_{16} \exp \left(-\left(\frac{|x_R- c_{17}|}{c_{18}}\right)^{c_{19}}\right) \biggr].
\end{eqnarray}
Finally, we allow for an additional scaling with $\sqrt{s}$, which is required to obtain the correct $\pi^\pm$ multiplicity at different energies. The functional form is given by
\begin{eqnarray}
  A(s) &=&  \frac{ 1+\left(\sqrt{s/c_{20}}\right)^{c_{21}-c_{22}}}{ 1+\left(\sqrt{s_0/c_{20}}\right)^{\,c_{21}-c_{22}}}\,
             \left(\sqrt{\frac{s}{s_0}}\right)^{c_{22}} 
  \label{eq:As}
\end{eqnarray}
In all these formulas, $p_T, \sqrt{s}$, the mass of the particles,
$\sqrt{s_0}$ and energies are intrinsically normalized to 1 GeV, in order to have dimensionless parameters. 

\subsection{$K^\pm$ model}
We define the  Lorenz invariant cross section by:
\begin{equation}
   \sigma_{\rm inv}= \sigma_0 (s) \,  d_1 \,F_K(s, p_T, x_R) \, A_K(s) 
   \label{eq:main_equation_K}
\end{equation}
with 
\begin{eqnarray}     
     \label{eq:function_prompt_K}
      &&\!\!F_K(s, p_T, x_R) = (1-x_R)^{d_2} 
        \exp(-d_3 \,p_T^{d_{4}} x_R) \, p_T^{d_5}  \\ \nonumber
      &&\;\;\times \exp\left[ -d_6 \sqrt{s/s_0}^{\;d_7} 
      \left(\sqrt{p_T^2 + m_K^2}-m_{K}\right)^{d_8 \sqrt{s/s_0}^{\;d_7}} \right] \;, 
\end{eqnarray}
where $m_{K}$ is the mass of the kaon and $d_i$ are the fit parameters reported in Table III of \us. The energy dependent normalization $A_K(s)$ is taken to be:
\begin{eqnarray}
  A_K(s) &=&  A_{K}^0\! (1-\frac{\sqrt{s_{\rm th}}}{\sqrt{s}} )\! (1+\sqrt{\frac{s}{d_9}}^{d_{10}-d_{11}}) \!
             \sqrt{s}^{d_{11}} \;\;\;
  \label{eq:As_K}
\end{eqnarray}
where $s_{\rm th}$ is the threshold energy for $K^+$ production and  $A_K^0$ is determined by the condition $A_K(s_0)=1$. 

\subsection{$K^0_S$ and $\Lambda$ model}
We define the  Lorenz invariant cross section of $K^0_S$ by:
\begin{equation}
   \sigma_{\rm inv}= \sigma_0 (s) \,  k_1 \,F_{K^0_S}(p_T, x_F) \, A_{K^0_S}(s),
   \label{eq:main_equation_k0s}
\end{equation}
with 
\begin{eqnarray}     
     \label{eq:k0s}
      &&\!\!F_{K^0_S}(p_T, x_F) = (1-|x_F|)^{k_2} 
        \times   \\ \nonumber
      &&\times \exp(-k_3 \,p_T^{k_{4}} |x_F|) \;\;p_T^{k_5} \exp\left[ -k_6 \;  p_T^{k_7} \right] \;, 
\end{eqnarray}
where $k_i$ are the fit parameters. The energy dependent normalization $A_{K^0_S}(s)$ is taken to be:
\begin{eqnarray}
  A_{K^0_S} (s) &=&  A_{K^0_S,0} \left(1-\sqrt{\frac{k_8}{s}}^{k_{9}-k_{10}} \right) \sqrt{s}^{k_{10}},
  \label{eq:As_K0s}
\end{eqnarray}
where the $A_{K^0_S,0}$ is determined by the condition $A_{K^0_S}(\sqrt{s_0}=17.3\rm{\,GeV})=1$ and the best-fit parameters are reported in Table VII of \us. The Lorentz-invariant cross section for $\Lambda$ follows the same functional form as that of $K^0_S$, differing only in the values of the fit parameters.

\end{document}